\documentclass[pra,aps,amsfonts,amsmath,twocolumn,showpacs,floatfix]{revtex4}
\usepackage{bm}
\usepackage{amsfonts}
\usepackage{amsmath}
\usepackage{graphicx}
\newcommand{\vek}[1]{\bm{\mathrm{#1}}}

\newcommand{\jv}{\vek{j}}

\newcommand{\nablav}{\vek{\nabla}}

\newcommand{\ev}{\vek{e}}

\newcommand{\pv}{\vek{p}}
\newcommand{\rv}{\vek{r}}
\newcommand{\Rv}{\vek{R}}
\newcommand{\Pv}{\vek{P}}
\newcommand{\vv}{\vek{v}}
\newcommand{\htil}{\tilde{h}}

\newcommand{\Ref}[1]{Ref.\@ \cite{#1}}
\newcommand{\Refs}[1]{Refs.\@ \cite{#1}}
\newcommand{\Fig}[1]{Fig.\@ \ref{#1}}

\newcommand{\Sec}[1]{Sec.\@ \ref{#1}}
\begin{document}
\title{Radial quadrupole and scissors modes in trapped Fermi gases
  across the BCS phase transition}
\author{Michael Urban}
\affiliation{Institut de Physique Nucl{\'e}aire, CNRS-IN2P3 and
Universit\'e Paris-Sud, 91406 Orsay Cedex, France}
\begin{abstract}
The excitation spectra of the radial quadrupole and scissors modes of
ultracold Fermi gases in elongated traps are studied across the BCS
superfluid-normal phase transition in the framework of a transport
theory for quasiparticles. In the limit of zero temperature, this
theory reproduces the results of superfluid hydrodynamics, while in
the opposite limit, above the critical temperature, it reduces to the
collisionless Vlasov equation. In the intermediate temperature range,
the excitation spectra have two or three broad peaks, respectively,
which are roughly situated at hydrodynamic and collisionless
frequencies, and whose strength is shifted from the hydrodynamic to
the collisionless modes with increasing temperature. By fitting the
time dependent quadrupole deformation with a damped oscillation of a
single frequency, we can understand the ``jump'' of the frequency of the
radial quadrupole mode as a function of interaction strength which has
recently been reported by the Innsbruck group.
\end{abstract}
\maketitle
%
\section{Introduction}
%
In the recent years, experiments with ultracold trapped fermionic
atoms like $^6$Li or $^{40}$K have become a powerful tool for
improving our understanding of strongly correlated Fermi systems. For
example, the continuous tunability of the interatomic interaction via
Feshbach resonances made it possible to study the crossover from a
Bose-Einstein condensate (BEC) of diatomic molecules to a BCS-type
superfluid. The creation of vortex lattices in rotating Fermi gases
with attractive interaction (scattering length $a < 0$) showed
unambiguously that the superfluid BCS phase was indeed reached
\cite{Zwierlein}.

By studying the collective modes of these systems, one can obtain a
lot of interesting information which cannot be obtained from static
properties alone. Near the BCS phase transition, several collective
modes of the trapped Fermi gases show interesting features like an
extremely strong damping or a sudden change of the frequency
\cite{KinastHemmer, Bartenstein, KinastTurlapov, Altmeyer,
Wright}. However, most of these experiments were done in the vicinity
of the Feshbach resonance ($k_F |a| \gtrsim 1$, where $k_F$ is the
Fermi momentum in the center of the trap), and in this strongly
interacting regime no theory is available so far which could
quantitatively explain the observed damping effects. An obvious
qualitative difference between the strongly and the weakly interacting
regimes is that in the latter ($k_F |a| < 1$) the normal phase is in
the collisionless regime, whereas in the former ($k_F |a| \gtrsim 1$)
collisions can produce a hydrodynamic behavior of the normal phase
\cite{Wright}.

The aim of the present paper is to discuss the properties of
experimentally accessible collective modes, namely of the radial
quadrupole and scissors modes, in the BCS regime ($k_F |a| < 1$). In
particular, we want to see how the properties of the modes change
across the superfluid-normal phase transition. At zero temperature,
the collective modes can be described within the framework of
superfluid hydrodynamics, but this is no longer true at non-zero
temperature, where superfluid and normal components coexist. In
\Ref{TaylorGriffin} it was proposed to use Landau's two-fluid hydrodynamics
in this case. However, although this approach may be valid in the
strongly interacting regime, it cannot be applied in the weakly
interacting regime because of the lack of collisions. The theoretical
framework which will be used in the present paper is the semiclassical
transport theory developed for clean superconductors by
Betbeder-Matibet and Nozi{\`e}res \cite{BetbederMatibet}, which has
recently been applied to the case of trapped Fermi gases
\cite{UrbanSchuck,Urban}. In the case of a quadrupole oscillation in a
spherical trap, this theory predicts a strong damping of the
hydrodynamic mode in the superfluid phase due to its coupling to the
collisionless normal component of the system, made of thermally
excited quasiparticles.

The article is organized as follows. \Sec{method} gives a brief
summary of the theoretical method (for details, see \Refs{UrbanSchuck}
and \cite{Urban}). In \Sec{quadrupole}, the results for the quadrupole
mode are discussed, and it is shown that the frequency jump observed
in a recent experiment at Innsbruck \cite{Altmeyer} can be
understood. In \Sec{scissors}, results for the scissors mode are
presented and \Sec{conclusions} contains further discussions and
conclusions.
\section{Theoretical method}
\label{method}
%
The method used in the present work to describe collective modes in
the BCS phase at finite temperature is based on the quasiparticle
transport theory by Betbeder-Matibet and Nozi\`eres
\cite{BetbederMatibet}. The basic degrees of freedom are the phase
$\phi(\rv,t)$ of the superfluid order parameter $\Delta(\rv,t)$,
describing the collective flow of the Cooper pairs, and the
quasiparticle distribution function $\nu(\rv,\pv,t)$, describing the
normal-fluid component. In equilibrium, one can choose $\phi = 0$, and
the quasiparticle distribution function $\nu$ reduces to the usual
equilibrium Fermi function $f(E)=1/(e^{E/T}+1)$, where $E =
\sqrt{h^2+\Delta^2}$ with $h = p^2/2m+V(\rv)-\mu$, $V$ and $\mu$ being
the trap potential and chemical potential, respectively. In the
non-equilibrium case, one writes $\Delta = |\Delta| e^{-2i\phi}$, such
that the collective velocity of the Cooper pairs is given by $\vv_s =
-\hbar\nablav\phi/m$. In this case, $\nu$ is defined in the rest frame
of the Cooper pairs, and the density $\rho$ and current $\jv$ are
given by
\begin{gather}
\rho = \int\!\frac{d^3 p}{(2\pi\hbar)^3}\, \Big(\frac{1}{2}-
  \frac{\htil_+}{2E_+}\big(1-2\nu_+\big)\Big)\,,\\
\jv = \rho\vv_s
  +\int\!\frac{d^3p}{(2\pi\hbar)^3}\,\frac{\pv}{m}\nu_-\,.
\end{gather}
The subscripts $+$ and $-$ denote the time-even and time-odd parts
of a given function (e.g., $\nu_\pm(\rv,\pv,t) =
(\nu(\rv,\pv,t)\pm\nu(\rv,-\pv,t))/2$),
\begin{equation}
\htil = \frac{(\pv-\hbar\nablav\phi)^2}{2m} + V - \hbar\dot{\phi} -
\mu
\end{equation}
is the hamitonian and
\begin{equation}
E = \sqrt{\htil_+^2+|\Delta|^2}+\htil_-
\end{equation}
the quasiparticle energy in the rest frame of the Cooper pairs.

The equation of motion of $\nu$ can be written in the compact form
\begin{equation}
\dot{\nu} = \{E,\nu\}\,,
\end{equation}
where $\{E,\nu\}$ denotes the Poisson bracket of $E$ and $\nu$,
similar to the usual Vlasov equation, which can be written as $\dot{f}
= \{h,f\}$, where $f$ denotes the usual distribution function. The
equation for $\phi$ can be obtained from the continuity equation
\begin{equation}
\dot{\rho}+\nablav\cdot\jv = 0\,,
\end{equation}
which becomes a second-order differential equation for $\phi$ if one
inserts the explicit expressions for $\rho$ and $\jv$.

In order to describe collective modes, one writes $V = V_0+V_1$, $\nu
= \nu_0 + \nu_1$, and $\phi = \phi_0+\phi_1$, with $\nu_0 = f(E_0)$
and $\phi_0 = 0$, the subscripts $0$ and $1$ referring to equilibrium
quantities and deviations from equilibrium, respectively. Assuming
that the perturbation $V_1$ is weak, one keeps only terms which are
linear in the deviations from equilibrium. The explicit equations of
motion for $\phi_1$ and $\nu_1$ are quite cumbersome and can be found
in \Refs{UrbanSchuck} and \cite{Urban}.

Note that, contrary to \Ref{Urban}, the mean-field (Hartree) potential
$g\rho$ ($g$ being the coupling constant) has been neglected in the
definitions of $h$ and $\htil$. If one wants to approach the crossover
region ($k_F |a| \sim 1$), the expression $g\rho$ for the mean-field
potential is not applicable. One possible solution of this problem is
to replace $g\rho$ by the real part of the single-particle self-energy
in ladder approximation \cite{Perali}. Anyway, previous experimental
and theoretical studies showed that the mean-field shift does not
modify the qualitative behaviour of the modes and leads only to minor
changes of the mode frequencies. For example, in the recent Innsbruck
experiment on the radial quadrupole mode \cite{Altmeyer}, the
frequency shift in the collisionless normal phase from $2\omega_r$
($\omega_r$ being the radial trap frequency) to $\approx 2.1\omega_r$
was attributed to the mean field. However, the mean field can
considerably change the equilibrium density profile, thereby modifying
the gap, the critical temperature, etc. For a quantitative study, the
inclusion of the mean field is therefore desirable, but beyond the
scope of the present paper.

Except in very simple cases (e.g., a sound wave propagating in a
uniform system \cite{UrbanSchuck}), the equations of motion for
$\nu_1$ and $\phi_1$ have to be solved numerically. In \Ref{Urban}, a
test-particle method was proposed which is similar to the one often
used for solving the Boltzmann equation. In the context of the
linearized theory, it is useful to write $\nu_1$ in the form
\begin{equation}
\nu_1(\rv,\pv,t) = \frac{df(E_0)}{dE_0} y(\rv,\pv,t)\,.
\end{equation}
The advantage of this rewriting is that $y$ is a smooth function in
phase space, contrary to $\nu_1$ [this is analogous to writing $f_1 =
f_0(1-f_0)\Phi$ in the context of the linearized Boltzmann equation,
which is a rather common trick]. The test-particle method consists in
replacing the continuous function $\nu_1(\rv,\pv,t)$ by a sum of
$\delta$ functions,
\begin{equation}
\nu_1(\rv,\pv,t) \to \sum_{i=1}^{N_\nu} y_i(t)\,
  \delta[\rv-\Rv_i(t)]\,\delta[\pv-\Pv_i(t)]
\end{equation}
where $y_i(t) = y[\Rv_i(t),\Pv_i(t),t]$. The trajectories of the test
particles in phase space, $\Rv_i(t)$ and $\Pv_i(t)$, satisfy the
following equations of motion:
\begin{equation}
\dot{\Rv} = \frac{\partial E_0}{\partial\Pv}\,,\qquad
\dot{\Pv} = -\frac{\partial E_0}{\partial\Rv}\,.
\end{equation}
These equations of motion do not conserve the energy $h_0$, but the
quasiparticle energy $E_0 = \sqrt{h_0^2+\Delta_0^2}$. This can lead to
surprising effects which make the numerical solution more involved
than that of usual Newtonian equations of motion. A well-known example
is Andreev reflection: If a particle ($h>0$) arrives at a point with
$\Delta = E$, it is reflected as a hole ($h<0$), and vice versa.

In the present calculations, the number of test particles is set to
$N_\nu = 10^5$ as in \Ref{Urban}. However, practically identical
results are already obtained with $N_\nu = 10^4$. In fact, the large
number of test particles was only needed in \Ref{Urban} in order to
obtain a stable mean field, while here, where the mean field is
neglected, a much smaller number of test particles is sufficient.

The time dependence of the coefficients $y_i(t)$ can rather easily be
solved. For the time dependence of the phase, $\phi$ is expanded in a
basis and the time dependence of the coefficients is determined by
minimizing the violation of the continuity equation. In the case of
the modes under consideration, the velocity field and therefore the
form of $\phi$ is well known. In this case, one or two basis functions
are completely sufficient \cite{Urban}.
%
\section{Radial quadrupole mode}
\label{quadrupole}
%
Let us start by discussing the radial quadrupole mode. As it is
usually the case in the present experiments, the unperturbed trap
potential is supposed to be cigar-shaped: $V_0(\rv) = \frac{1}{2} m
[\omega_r^2 (x^2+y^2)+\omega_z^2 z^2]$ with $\omega_r \gg
\omega_z$. The radial quadrupole mode is excited at time $t = 0$ by
applying a perturbation of the form $V_1(\rv,t) = \hat{V}_1(\rv)
\delta(t)$ with $\hat{V}_1(\rv) \propto x^2-y^2$. The equations of
motion are then solved for $t > 0$. The excitation spectrum is
obtained as the Fourier transform of the time evolution of the
quadrupole moment $Q = \langle x^2-y^2\rangle$.

We choose here parameters similar to the recent Innsbruck experiments:
$N = 4\times 10^5$ atoms, trap frequencies $\omega_r = 2\pi\times 370$
Hz, $\omega_z = 2\pi\times 22$ Hz. Since on the one hand our theory is
only valid in the BCS limit, but on the other hand the scattering
length of $^6$Li saturates at quite a large value at high magnetic
fields, we choose as a compromise $1/k_F a = -1.5$. For these
parameters, the critical temperature within BCS theory is $T_C \approx
0.058 T_F \approx 43$ nK (where $T_F = \epsilon_F/k_B$ is the Fermi
temperature). However, one should remember that BCS theory cannot be
trusted quantitatively. To give an example, the inclusion of the
correlated density according to Nozi\`eres and Schmitt-Rink
\cite{Nozieres}, keeping the number of atoms $N=4\times 10^5$ and the
parameter $1/k_F a = -1.5$ fixed, reduces the critical temperature by
almost $20\,\%$ to $T_C \approx 0.049 T_F$ ($k_F$ and $T_F$ being
defined as usual by the density of a non-interacting system with the
same number of atoms).

\begin{figure}
\includegraphics[width=6.9cm]{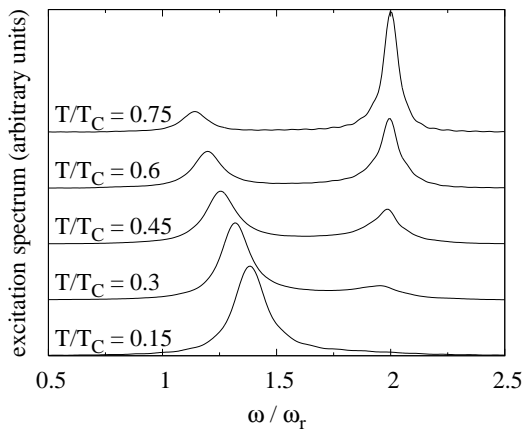}
\caption{Excitation spectra as function of the excitation frequency
$\omega$ (in units of the radial trap frequency $\omega_r$) for
different temperatures ranging from $0.15 T_C$ to $0.75 T_C$ (from
bottom to top). For a better visibility of the strength contained in
the narrow peaks at $2\omega_r$, all spectra have been folded with a
Lorentzian of width $\approx 0.03 \omega_r$.}
\label{fig1}
\end{figure}

The resulting excitation spectra for different temperatures $T$ are
shown in \Fig{fig1}.
At $T = 0.15 T_C$, the spectrum is dominated by a peak situated at the
hydrodynamic frequency, $\sqrt{2} \omega_r$. There is, however, a
considerable broadening of the peak, i.e., damping, due to the
coupling of the hydrodynamic mode to the thermally excited
quasiparticles. At higher temperatures, the peak is further broadened
and slightly shifted towards lower frequencies. At the same time, a
second peak appears, which is situated at $2\omega_r$, the frequency
of the radial quadrupole mode in an ideal (collisionless normal-fluid)
Fermi gas. Close to $T_C$, this mode is practically undamped (the
width of the peak at $2\omega_r$ in the uppermost spectrum in
\Fig{fig1} is due to an artificial broadening which has been
introduced in order to improve the visibility of the peaks). As one
can clearly see, the peak position does not move from
$\sqrt{2}\omega_r$ to $2\omega_r$ with increasing temperature, but the
strength is shifted from the lower to the upper peak. There seems to
be no simple explanation for the downshift of the hydrodynamic mode,
but it can partially be understood as a consequence of the usual
repulsion between two coupled modes. The general behavior is very
similar to that found in previous work on the quadrupole mode in
spherical traps \cite{BruunMottelson,GrassoKhan,Urban}.

This finding suggests the following interpretation of the results of
the Innsbruck experiment \cite{Altmeyer}, where a sudden ``jump'' of
the frequency as a function of the interaction strength was found. In
the analysis of the experiment, the deformation of the atom cloud was
fitted with an ansatz function containing only a single frequency and
damping rate. Let us try what happens if we apply the same analysis to
our theoretical results.

In order to follow as closely as possible the experiment, we will do
the calculation as a function of the coupling strength for fixed
temperature, which we choose rather arbitrarily as $T = 0.046 T_F$. We
also simulate the experimental initial conditions, which are different
from the $\delta$-like perturbation used above. In the experiment, the
potential is first slowly deformed from $V_0(\rv)$ to
$V_0(\rv)+V_1(\rv)$, with $V_1(\rv)\propto x^2-y^2$, giving enough
time to the collisions to keep the momentum distribution spherical,
and then the deformation $V_1(\rv)$ is suddenly switched off at $t =
0$. Unfortunately there is one step in the experiment which cannot be
simulated within the present linearized theory, namely the expansion
of the cloud before the picture is taken. However, if one assumes that
the picture taken after the expansion reflects in some way the state
of the system at the moment when the trap was switched off, it seems
plausible to assume that the expansion is not important for the
observed frequency and damping, but only for the observed amplitude
and phase of the oscillation.

\begin{figure}
\includegraphics[width=6.9cm]{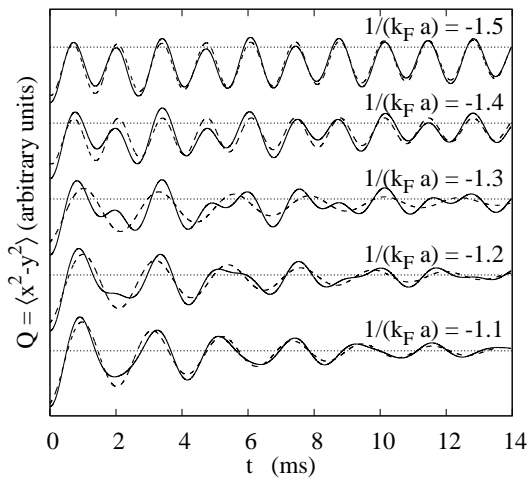}
\caption{Quadrupole moment as function of time for different values of
$1/k_F a$ ranging from $-1.1$ to $-1.5$ (from bottom to top) for
fixed temperature $T$.}
\label{fig2}
\end{figure}

In \Fig{fig2} the obtained time dependence of the quadrupole moment
$Q(t)$ is shown for different values of $1/k_F a$. The solid lines
are the calculations, while the dashed lines correspond to fits with
functions of the form
\begin{equation}
Q_\mathit{fit}(t) = A e^{-\kappa t} \cos(\omega
t+\phi)+C e^{-\xi t}\,.
\end{equation}
The parameters $A$, $\kappa$, $\omega$, $\phi$, $C$, and $\xi$ are
determined by minimizing the error
\begin{equation}
\langle(\Delta Q)^2\rangle =
  \frac{1}{t_\mathit{max}}\int_0^{t_\mathit{max}} dt\,
    [Q(t)-Q_\mathit{fit}(t)]^2\,,
\end{equation}
where $t_\mathit{max} = 14$ ms is chosen as in \Ref{Altmeyer}. It can
be seen that for $1/k_F a = -1.1$ and for $1/k_F a = -1.5$ the fit
reproduces almost exactly the full calculation, while for the
intermediate couplings, in particular for $1/k_F a = -1.3$, the fit
is quite bad.

\begin{figure}
\includegraphics[width=6.9cm]{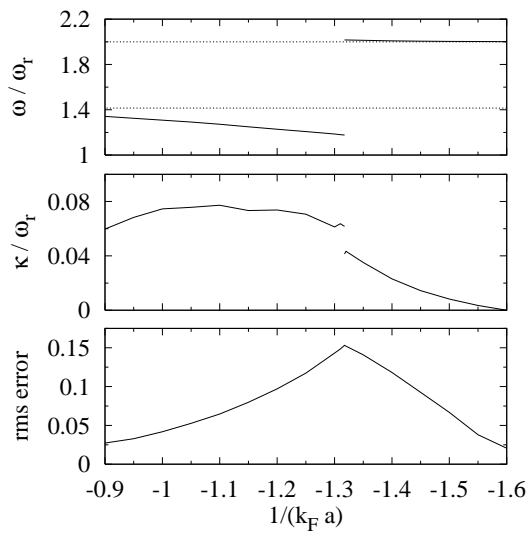}
\caption{From top to bottom: Fitted frequency $\omega$, damping rate
$\kappa$, and corresponding rms error of the fit (definition see text)
for the radial quadrupole mode as functions of $1/k_F a$ for fixed
temperature $T$. The dotted lines in the upper panel indicate the
prediction of superfluid hydrodynamics ($\sqrt{2}\omega_r$) and the
prediction for an ideal Fermi gas ($2\omega_r$).}
\label{fig3}
\end{figure}

In \Fig{fig3} the fitted frequency $\omega$ and damping rate $\kappa$
are shown as functions of the parameter $1/k_F a$. The third panel
shows the corresponding rms error of the fit, defined as
$\sqrt{\langle (\Delta Q)^2\rangle}/|Q(t=0)|$. There is a couple of
striking similarities with the experimental results: At strong
coupling, where $T_C$ is large and hence $T/T_C$ is small, the fit
gives the hydrodynamic frequency. With decreasing coupling, i.e.,
increasing $T/T_C$, the damping increases and the frequency of the
hydrodynamic mode is shifted downwards. The quality of the fit becomes
quite bad (this corresponds probably to the shaded area in
\Ref{Altmeyer}), and at some critical coupling strength the fit jumps
to the collisionless (normal-fluid) mode. As seen in the experiment,
the damping rate of the collisionless mode is weaker than that of the
hydrodynamic mode before the jump.

It should be noticed that with the present choice of parameters, the
jump happens at $1/k_F a\approx -1.32$, corresponding to $T/T_C
\approx 0.6$. This means that in the center of the trap the system is
still superfluid, but since this region contributes only very little
to the quadrupole moment, the quadrupole mode behaves almost as if the
system was in the collisionless normal-fluid phase. 

In spite of the surprising qualitative agreement with the findings of
\Ref{Altmeyer}, there is no exact agreement on a quantitative
level. First of all, in the experiment the sudden transition from
hydrodynamic to collisionless normal-fluid behavior happens already at
a stronger coupling than in the calculation. The position of the jump
depends on the temperature, and if the temperature in the experiment
had been higher than ours ($0.046 T_F$), this could explain the
difference. In addition, one should remember that the critical
temperature predicted by BCS theory is too high. The second problem is
the damping rate: The experimental damping rates are larger than the
theoretical ones by roughly a factor of 2 near the jump, and even
worse in the case of the collisionless normal-fluid mode far away from
the transition. In fact, the experiment is not in the collisionless
limit and collisions still play a role. In order to improve the theory
in this respect, one would have to include a collision term similar to
that of the Boltzmann equation into the quasiparticle transport
equation. The small anharmonicity of the experimental trap potential
could be an additional source of damping. However, a calculation where
the radial harmonic potential, $\frac{1}{2}m\omega_r^2(x^2+y^2)$, was
replaced by a Gaussian one, $-U_0 e^{-(x^2+y^2)/d^2}$, with a
realistic width $d = 0.05$ mm, showed that this effect can explain
only $\sim 10\%$ of the experimentally observed damping in the normal
phase. The third point concerns the frequencies: On the one hand,
although the theory gives a down-shift of the hydrodynamic mode before
the transition, it is not as strong as the down-shift observed in the
experiment. On the other hand, the theoretical frequency of the
collisionless normal-fluid mode above the transition is too low,
because the mean-field has been neglected.
%
\section{Scissors mode}
\label{scissors}
%
In the case of the radial quadrupole mode the velocity field is always
irrotational and it does not change significantly from the superfluid
to the normal-fluid phase. The different frequencies are a consequence
of the Fermi surface distortion during the oscillation in the
collisionless normal-fluid phase. This is different from the
``scissors mode,'' which has also been studied in a recent experiment
\cite{Wright}. In this experiment, the trap was deformed in the $xy$
plane: $V_0(\rv) = \frac{1}{2} m [\omega_x^2 x^2 + \omega_y^2 y^2
+\omega_z^2 z^2]$, with $\omega_x > \omega_y \gg \omega_z$. The
scissors mode corresponds to a motion where the deformation of the
atom cloud in the $xy$ plane is not aligned with that of the trap, so
that the angle between them oscillates. In the normal-fluid phase,
such a motion can be realized by rotating the gas back and forth
around the $z$ axis [velocity field $\vv(\rv)\propto \ev_z\times
\rv$], but also by an irrotational motion [velocity field
$\vv(\rv)\propto \nablav(xy)$]. In an ideal gas, these two kinds of
motion correspond to two distinct modes with frequencies $\omega_-$
and $\omega_+$, respectively, given by $\omega_\pm = \omega_x \pm
\omega_y$. In the superfluid phase, of course, the rotational mode is
suppressed and only the irrotational one exists. In the
zero-temperature limit, its frequency approaches the hydrodynamic
value $\omega_\mathit{sc} = \sqrt{\omega_x^2+\omega_y^2}$
\cite{GueryOdelin}.

\begin{figure}
\includegraphics[width=6.9cm]{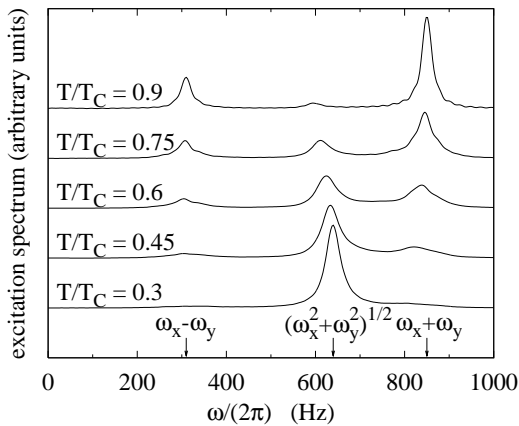}
\caption{Excitation spectra of the scissors mode as functions of the
excitation frequency $\omega$ for different temperatures ranging from
$0.3 T_C$ to $0.9 T_C$ (from bottom to top). The spectra have been
folded with a Lorentzian of width $\approx 2\pi\times 10$ Hz. The
arrows indicate the predicted frequencies in the hydrodynamic
($\sqrt{\omega_x^2+\omega_y^2}$) and collisionless limit
($\omega_x\pm\omega_y$), respectively}
\label{fig4}
\end{figure}

The excitation of the scissors mode consists in turning the trap
potential around the $z$ axis by a small angle. In the small-amplitude
limit, this corresponds to an excitation operator $\hat{V}_1(\rv)
\propto xy$. The excitation spectra of the scissors mode for different
temperatures, which are obtained analogously to those of the radial
quadrupole mode (i.e. Fourier transform of $\langle xy\rangle$ after a
$\delta$-like perturbation), are shown in \Fig{fig4}
for the case of $N = 3\times 10^5$ atoms in a trap with $\omega_x =
2\pi\times 580$ Hz, $\omega_y = 2\pi\times 270$ Hz, $\omega_z =
2\pi\times 22$ Hz, and $1/k_F a = -1.5$. At $T = 0.3 T_C$, the
spectrum has a single broad peak at the hydrodynamic frequency. The
damping rate (width) is comparable with that of the hydrodynamic
radial quadrupole mode at $T = 0.3 T_C$. With increasing temperature,
the two collisionless normal-fluid modes appear at the predicted
frequencies, while the strength of the hydrodynamic mode decreases and
its frequency is slightly shifted downwards. At $T = 0.6 T_C$ and
$0.75 T_C$ the scissors-mode spectrum has three pronounced
peaks. There is a striking analogy between these results and the
experimental observations concerning the scissors mode in a BEC with a
thermal cloud, where the BEC oscillates with $\omega_\mathit{sc}$
which is slightly shifted downwards while the thermal cloud oscillates
with $\omega_\pm$ \cite{Marago}.

The results of \Fig{fig4} are quite different from those found in the
experiment \cite{Wright}. However, this is not surprising, since the
experiment was not done in the BCS limit but for $1/k_F a = -0.45$,
where in particular the collisions are very important. As one can see
in Fig.\@ 2 of \Ref{Wright}, the phase transition in the experiment
does not lead directly from the superfluid to the collisionless
regime, but the system enters first a collisionally hydrodynamic
regime (i.e., the mode frequency does not change) and reaches the
collisionless phase only at much higher temperature. As shown in
\Ref{BruunSmith}, the transition from collisional hydrodynamic to the
collisionless regime leads to a continuous increase of the upper
(irrotational) branch of the scissors mode. This explains why in the
experiment no sign of a sudden jump of the scissors mode frequency has
been observed. It would be interesting to see if in the BCS regime,
e.g. at $1/k_F a = -1.5$, the experiment would observe a similar
jump as in the case of the quadrupole mode.
%
\section{Conclusions}
\label{conclusions}
In the present paper, the behavior of the radial quadrupole and of the
scissors mode across the BCS phase transition was theoretically
studied in the framework of a semiclassical theory, in which
superfluid hydrodynamics is coupled to a quasiparticle transport
theory for the normal component. The theory proved to be applicable to
realistic situations (large numbers of atoms, arbitrary trap
geometry), and it provides an explanation of the surprising behavior
of the radial quadrupole mode as a function of $1/k_F a$ as observed
by the Innsbruck group \cite{Altmeyer}: The damping near the BCS phase
transition is due to the coupling between the hydrodynamic mode and
the thermally excited quasiparticles. In addition to the damping, this
coupling leads to a two-peak spectrum, the two peaks belonging,
respectively, to the hydrodynamic mode of the superfluid component and
the collisionless mode of the normal-fluid component. The sudden jump
from the hydrodynamic to the collisionless frequency reported in
\Ref{Altmeyer} is probably a consequence of the fit with a single
frequency in the analysis of the experiment.

It should be mentioned that an alternative explanation of the damping
was suggested in \Ref{CombescotLeyronas}: The frequency shift and
damping are not necessarily a signal for a superfluid-normal phase
transition. They could also be related to the breakdown of superfluid
hydrodynamics when $\hbar\omega$ is of the same order of magnitude as
the pairing gap $\Delta$. In \Ref{GrassoKhan}, effects of this kind
were quantitatively studied by using a fully quantum mechanical
formalism [quasiparticle random phase approximation, (QRPA),
corresponding to the time-dependent Bogoliubov-de Gennes equations in
the case of small deviations from equilibrium], and it was shown that
indeed, even at zero temperature, considerable damping and deviations
from the hydrodynamic frequency can occur if $\hbar\omega/\Delta$ is
not small. Although such effects may also play a role in the recent
experiments (maybe they can give rise to an additional downshift of
the hydrodynamic quadrupole mode, which would help to quantitatively
explain the data), they are probably less important than the coupling
between the hydrodynamic mode and the thermally excited quasiparticles
considered here: In \Ref{Urban} it was shown that for a ratio of mode
frequency to zero-temperature gap of $\hbar\omega/\Delta(T=0) \sim
0.3$ (which is similar to the most unfavorable ratio of all the cases
considered in the present paper), the results of the semiclassical
theory and of the quantum mechanical QRPA calculation for the
quadrupole mode in a spherical trap are in good agreement, even if the
finite-temperature gap is smaller.

Possible extensions of this theory which are needed for giving more
quantitative results at the limit between the BCS and the crossover
regime ($k_F |a|\sim 1$) have been mentioned. These are, on the one
hand, the inclusion of quasi-particle collisions in order to reproduce
the observed damping, and on the other hand the inclusion of the
single-particle self-energy in order to account for the ``mean field''
energy shift.

However, these extensions may not be enough to give a complete
description of the BCS-BEC crossover ($k_F |a| \gtrsim 1$), including
the unitary limit ($k_F |a|\to \infty$). In this regime, the
normal-fluid component of the system is not only made of thermally
excited quasiparticles, which are included in the present theory, but
also of pair fluctuations \cite{Ohashi}, which persist even above the
critical temperature $T_C$ up to the ``pair breaking temperature''
$T^*$ and which cannot easily be accounted for in the present
theoretical framework.
\acknowledgments I thank G.M. Bruun and E.R. S\'anchez Guajardo for
  discussions, comments and suggestions.

\end{document}